\documentstyle[12pt]{article}
\newcommand{\bm}[1]{\mbox{\boldmath $#1$}}
\begin{document}
\title{
Deformed Hartree-Fock Calculation of Proton-Rich
Nuclei\footnote{
To be published in a special issue of Nuclear Physics {\bf A} for the
proceedings of the international symposium on physics of unstable
nuclei (Niigata, Japan, October 31 - November 3, 1994) edited by
H.\ Horiuchi, K.\ Ikeda, K.\ Sato, Y.\ Suzuki, and I.\ Tanihata}
}
\author{
N.~Tajima, N.~Onishi, and S.~Takahara\\
Institute of Physics, College of Arts and Sciences,\\
University of Tokyo, Komaba, Meguro-ku, Tokyo, 153, Japan
}
\date{November 1, 1994}
\maketitle

\begin{abstract}
We perform Hartree-Fock+BCS calculations for even-even nuclei with $2
\leq Z \leq 82$ and $N$ ranging from outside the proton drip line to
the experimental frontier on the neutron-rich side.  The ground state
solutions are obtained for 737 nuclei, together with
shape-coexistence solutions for 480 nuclei. Our method features the
Cartesian-mesh representation of single-particle wavefunctions, which
is advantageous in treating nucleon skins and exotic shapes.  The
results are compared with those of the finite-range droplet model of
M{\o}ller et al.\ as well as the experimental values.
\end{abstract}

\section{INTRODUCTION} \label{s_intro}


Several theoretical frameworks have been presented for the sake of
global (on the $N$-$Z$ plane) calculations of nuclear ground state
properties.  One of the most elaborate results is given by M{\o}ller
et al.\ \cite{MNM94} using the finite-range droplet model with a
microscopic shell correction (FRDM). Another extensive calculation is
carried out by Aboussir et al.\ \cite{APD92} in the extended
Thomas-Fermi approach.  The former as well as the latter methods can
be regarded as approximations to the Hartree-Fock (HF) equation.
While mean-field calculations under spherical symmetry are very easily
done nowadays, those including deformation still requires long
computation time for global calculations.


In this paper we show the results of our global HF+BCS calculations
and compare them with those of FRDM and experiment.  Most of the
methods to treat deformed nuclei\cite{MNM94,APD92,DG80} express
single-particle wavefunctions by expansion in a harmonic oscillator
basis.  In contrast, we use a HF+BCS code {\em ev8}\cite{BFH85}, which
features a Cartesian-mesh representation of single-particle
wavefunctions.  The mesh representation is expected to be suitable for
describing the atomic nucleus because the saturation property of
nuclear matter guarantees the suppression of large-momentum components
in wavefunctions.

The advantage of the mesh representation for the study of unstable
nuclei is the capability to treat nucleon skins and halos, while they
cannot be described correctly in the oscillator basis expansion since
the asymptotic form of wavefunctions far from the nuclear surface is
determined by the basis.

Another advantage of the mesh representation is that it can treat
exotic (e.g. high-multipole) shapes and large (e.g. super and hyper)
deformations without preparing a special basis for each shape.

\section{METHOD OF CALCULATION} \label{s_method}


In the code {\em ev8}, an octant of a nucleus is placed in a corner of
a box (13 $\times$ 13 $\times$ 14 fm$^3$), imposing a symmetry with
respect to reflections in $x$-$y$, $y$-$z$, and $z$-$x$ planes (the
point group $D_{2h}$).  The mesh size takes on 1 fm.  The correction
for the inaccuracy due to this finite mesh size is necessary only for
the total binding energy\cite{BFH85}. (We need a precision of 0.1 MeV
in the total kinetic energy of 4 GeV for $^{208}$Pb.)  The correction
to be added is,
\begin{equation}
  \Delta E(Z,N) = 0.00595 \cdot (Z+N) + 0.1090 \cdot (N-Z) + 1.24
\;\;\;\; \mbox{[MeV]},
\end{equation}
which was obtained by fitting to the outputs of a spherical HF+BCS
code\cite{Re91} with sufficiently small grid size (0.2 fm).
In Fig.\ \ref{f_mesh}, the nucleon density of a nucleus $^{138}$Dy in
the $x$-$z$ plane is shown as an illustration of our mesh
representation.


The code uses the Skyrme interaction.  Among the widely used
parameters of the interaction, we choose the SIII\cite{BFG75}.  Its
validity has been examined in many nuclear structure calculations.  In
particular, its single-particle spectrum is in good agreement with
experiment.  It also reproduces fairly well the $N-Z$ dependence of
the binding energy compared with other widely-used parameter
sets\cite{TBF93}. On the other hand, its incompressibility is said to
be too large.


The pairing correlation plays an important role in determining the
nucleon drip lines\cite{DFT84}.  It also influences the deformation,
sometimes strongly.  For example, one can see in ref.\ \cite{TFB93}
that the potential energy curve of $^{186}$Pb is substantially changed
when the pairing interaction strength is slightly reduced.  Therefore,
it should be kept in mind that more accurate treatment of the pairing
is necessary for deformed and/or near-drip-line nuclei than for
spherical stable nuclei.

The BCS method cannot be applied to nuclei near the neutron drip line
because the Fermi energy is so high that positive energy HF orbitals,
whose wavefunctions are extended over the box, are occupied through
pair-scattering processes.  This brings about an unphysical situation
in which neutron gas surrounds the nucleus.  This problem is resolved
in the Hartree-Fock-Bogolyubov (HFB) theory, in which ordinary and
pair densities are localized when the Fermi level is
negative\cite{DFT84}.

On the other hand, for nuclei near the proton drip line, the HF+BCS
method is still useful to obtain localized solutions because the
Coulomb barrier keeps the wavefunctions of positive energy HF orbitals
localized spatially.  In practice, to prevent the tunneling through
the barrier, we modify the Coulomb potential outside the barrier so
that it is higher than the cut-off energy
($\lambda_{\rm HF}^{\tau}+7.3$MeV) of the pairing interaction.


We use a seniority pairing force, whose pair-scattering matrix
elements are defined as a constant multiplied by cut-off factors
depending on the single-particle energy $\epsilon_i$:
\begin{eqnarray}
  \langle i \bar{\imath} | V_{\rm pair}^{\tau} | j \bar{\jmath} \rangle
  & = & - G_{\tau} f_{\tau}(\epsilon_i) f_{\tau}(\epsilon_j), \\
  f_{\tau}(\epsilon) & = & \Biggl\{ 1 + \exp \frac{\epsilon -
  \lambda_{\rm HF}^{\tau} - 5 \; {\rm MeV}}
  {0.5 \; {\rm MeV}} \Biggr\}^{-1/2}
  \theta(\lambda_{\rm HF}^{\tau} + 7.3 \; {\rm MeV} - \epsilon),
  \label{eq_f}
\end{eqnarray}
where $\tau$ means proton or neutron while $\lambda_{\rm HF}^{\tau}$
is the Fermi level of the HF (normal) state.
For neutrons, the righthand side of eq.\ (\ref{eq_f}) is
multiplied furthermore by $\theta(-\epsilon)$.
We need a prescription to determine the strength $G_{\tau}$ for each
nucleus in the entire region of the nuclear chart.  For this purpose,
we develop a method based on the continuous spectrum approximation
using the Thomas-Fermi single-particle level density.  As the average
pairing gap, we adopt the classical empirical formula $\Delta$ = 12
MeV/${\sqrt{A}}$.


The imaginary-time time-dependent-Hartree-Fock evolution
method\cite{DFK80} is used to obtain the solutions to the HF+BCS
equation. Usually, 500-2000 steps are necessary to obtain a solution
for each nucleus, where the time step is chosen as
$1.5 \times 10^{-24}$ sec.

In order to search a prolate (an oblate) solution, we first exert an
external potential proportional to $Q_{\rm z}$ on the initial
wavefunction until its quadrupole deformation parameter,
\begin{equation}
  \delta_2 = 3\langle Q_{\rm z}\rangle / 4\langle r^2 \rangle,
\end{equation}
satisfies $\delta_2 > 0.1$ ($\delta_2 < -0.1$).  Then, we switch off
the external potential, let the wavefunction evolve by itself, and see
if it converges to a deformed local minimum.

\section{RESULTS}


We have performed the HF+BCS calculations with the Skyrme SIII force
for 752 even-even nuclei with $2 \le Z \le 82$ and, for each $Z$, $N$
ranging from several units less than the proton drip line to the
heaviest isotope observed.  The extension of the calculation to the
neutron drip line will be deferred until a deformed HFB method is
developed.  Spatially localized solutions are obtained for 737 nuclei.
Note that some of them are outside the proton drip line: The
localization is due to the lifting of the Coulomb potential outside
the barrier.  We have also searched for prolate and oblate
shape-coexistence solutions and found the second minimum for 480
nuclei.

Although our method allows triaxial shapes, all the solutions we
obtained are axially symmetric: The $Y_{2,2}$ as well as the $Y_{4,2}$
deformations are almost vanishing, while the $Y_{4,4}$ deformation is
larger but very small.  It is possible, however, that local minima
exist at $\gamma \not= 0^{\circ}$ points in the $\beta$-$\gamma$ plane
which are separated from the axial path by a potential barrier.

Among the quantities which can be extracted from our results are the
binding energy, even-multipole deformation parameters, radius, skin
thickness (halo is not found), single-particle levels, and pairing
gaps.  We also have the information on the prolate-oblate difference
of these quantities.

In Fig. \ref{f_def}, the quadrupole deformation parameter $| \delta_2
|$ is shown for the ground states of nuclei with $42 \le Z \le 58$.
One can see the development of deformation with the increase of the
number of valence particles and holes.  One can also see that oblate
minima are competing with prolate ones for not very large deformations
($| \delta_2 | < 0.2$).  By comparing Fig. \ref{f_def} with the result
of FRDM\cite{MNM94}, we find conspicuous differences for $40 < Z,N <
50$, where only the latter predicts very large deformations.

In Fig. \ref{f_s2p}, two-proton separation energies are plotted for
nuclei in the neighbourhood of $^{100}$Sn.  The results of
FRDM\cite{MNM94} and experimental values\cite{Au89} are also shown.
The agreement between them looks more impressive considering that the
SIII force was made in 1975 while FRDM was revised in 1993.  A
noticeable discrepancy is that the jump across the $Z=50$ major shell
closure is smaller in experiment than our calculation.

In Fig. \ref{f_skin}, the thickness of the proton skin is shown. The
definition of the skin is taken from ref. \cite{FOT93}: The condition
for a point $\bm{r}$ to be in the proton skin is,
\begin{equation}
  \rho_{\rm p}(\bm{r}) > 4 \rho_{\rm n} ( \bm{r} )
  \;\;\;\; \mbox{and} \;\;\;\;
  \rho_{\rm p}(\bm{r}) > \rho_{\rm p} ( 0 ) / 100.
\end{equation}
When the nucleus is deformed, the thickness of the skin can depend on
the direction. In the figure the skin thickness in the $z$-axis
(symmetry axis) is plotted. The skin in the $x$-$y$ (equatorial) plane
is usually slightly thinner for prolate shapes because of the Coulomb
repulsion.  One can see that the proton skin is very thin ($<$0.3 fm)
for nuclei inside the proton drip line except some light ($Z \le 20$)
nuclei.  For all the nuclei with $50 < Z \le 82$, which are not shown
in the figure, the proton skin does not exist.


\vspace{\baselineskip}

We thank Dr.~P.~Bonche, Dr.~H.~Flocard, and Dr.~P.-H.~Heenen for
providing the Hartree-Fock+BCS code {\em ev8} on a Cartesian mesh.
We are grateful to Dr.~Flocard for useful discussions, too.
The computation was financially supported by RCNP, Osaka University,
as RCNP Computational Nuclear Physics Project (No.~94-B-01).


\newpage

\noindent {\bf FIGURE CAPTIONS}

\newcounter{figno}
\begin{list}
{Fig. \arabic{figno}. }{\usecounter{figno}
   \setlength{\labelwidth}{1.3cm}
   \setlength{\labelsep}{0.5mm}
   \setlength{\leftmargin}{8.5mm}
   \setlength{\rightmargin}{0mm}
   \setlength{\listparindent}{0mm}
   \setlength{\parsep}{0mm}
   \setlength{\itemsep}{0.5cm}
   \setlength{\topsep}{0.5cm}
}
\item \label{f_mesh}
The nucleon density of $^{138}$Dy in the $x$-$z$ plane calculated with
the HF+BCS method with the SIII force.  This nucleus is located just
inside the proton drip line and has a prolate shape ($\delta_2=0.315$)
which is symmetric in the $z$-axis. Dashed lines correspond to the 1
fm mesh.
\item \label{f_def}
The quadrupole deformation parameter $| \delta_2 |$ as a function of
the neutron number $N$ for some isotope chains near Sn calculated with
the HF+BCS method with the SIII force.  On the lefthand (righthand)
side, $Z=42,44,46,48$ ($Z=58,56,54,52$) isotopes are designated by
triangles connected with solid lines, squares with short-dash lines,
diamond shapes with long-dash lines, and circles with dot-dash lines,
respectively.  The solid symbols are used for prolate states, while
open symbols are used for oblate ones.  All the $Z=50$ isotopes are
spherical and have no deformed local minima.
\item \label{f_s2p}
Two-proton separation energy in the neighborhood of $^{100}$Sn.  The
calculation with the HF+BCS method with the SIII force is designated
by solid circles. The solid lines connect the solid circles of the
same neutron number.  The neutron number is printed at an end of each
line.  The results of FRDM\cite{MNM94} are represented by open
circles, while the experimental values\cite{Au89} are designated by
plus marks.
\item \label{f_skin}
The thickness of the proton skin in the $z$-axis (the symmetry axis)
calculated with the HF+BCS method with the SIII force.  The definition
of the skin is given in the text.  The solid (open) circles are used
for nuclei with $S_{\rm 2p}>0$ ($S_{\rm 2p}<0$).  The lines connect
isotones.  The number printed near each line means the neutron number
of the isotone chain.
\end{list}

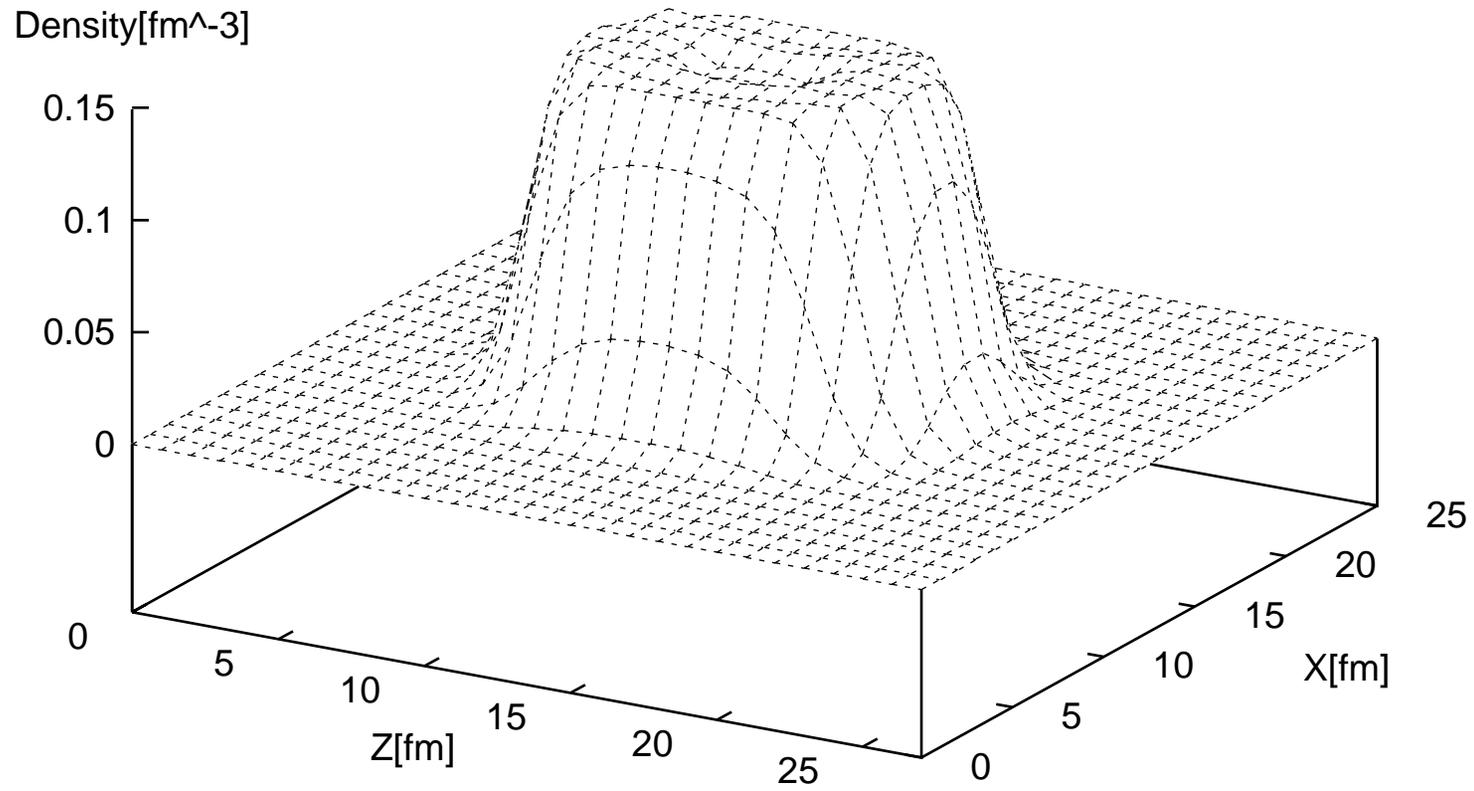
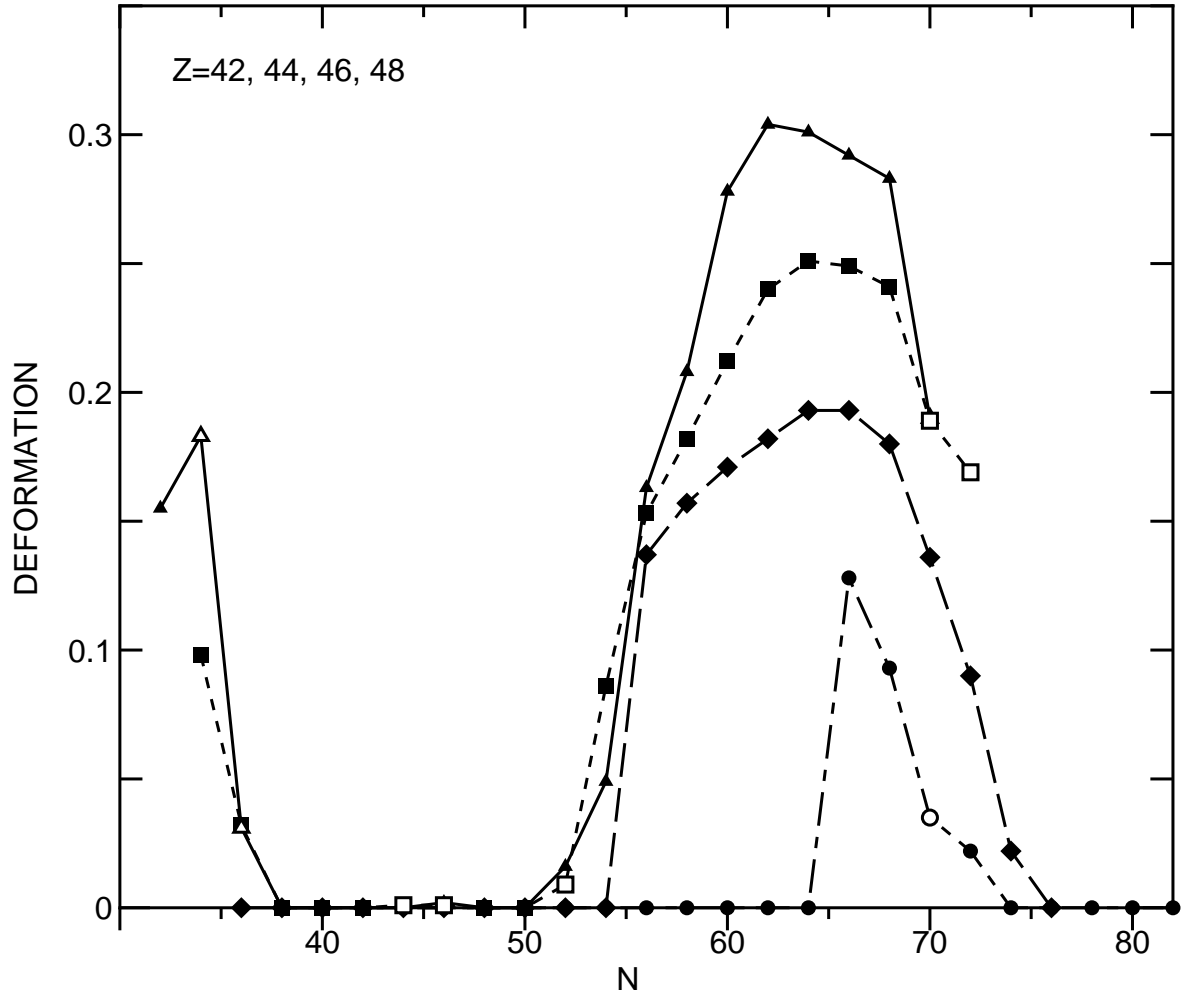
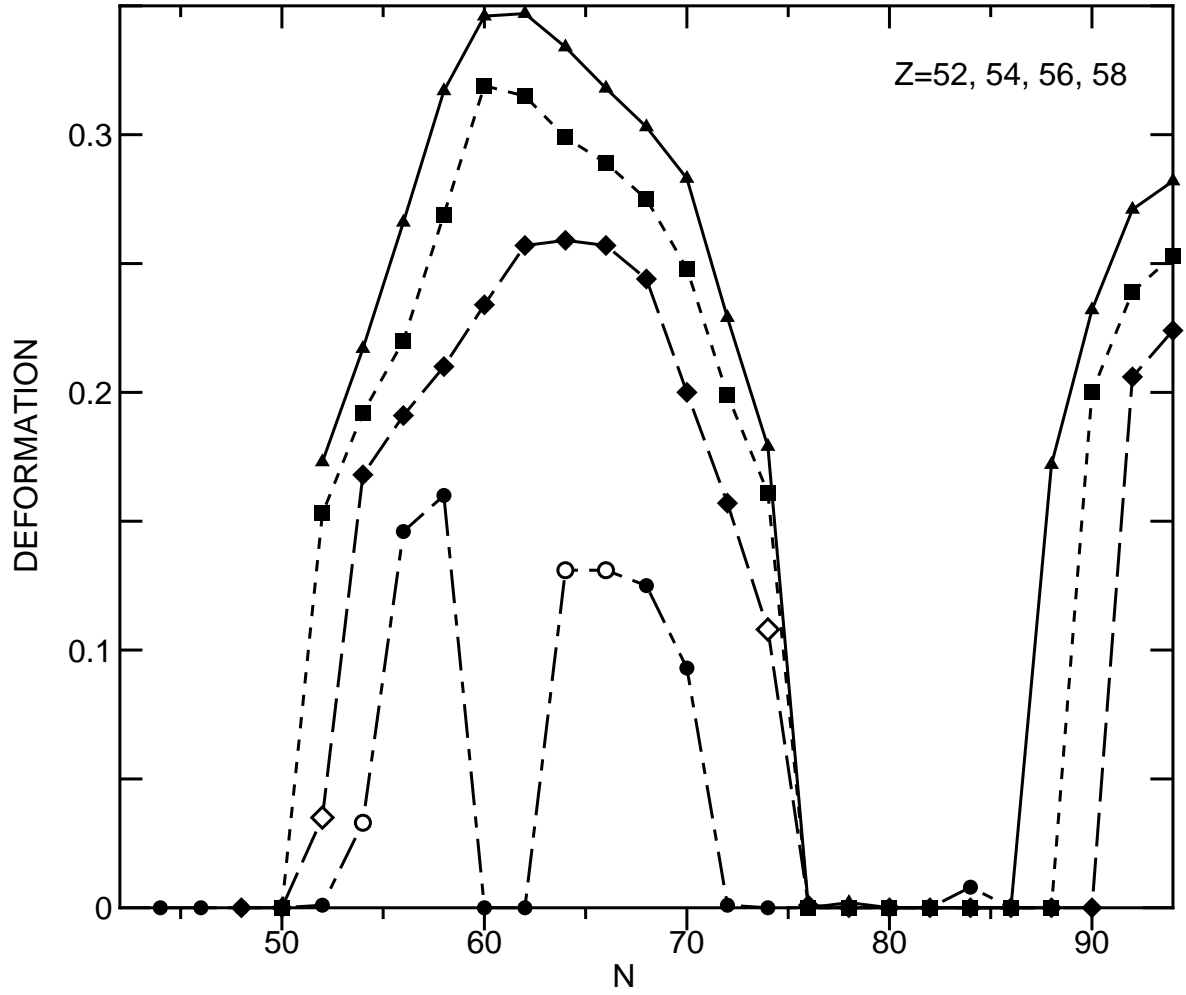
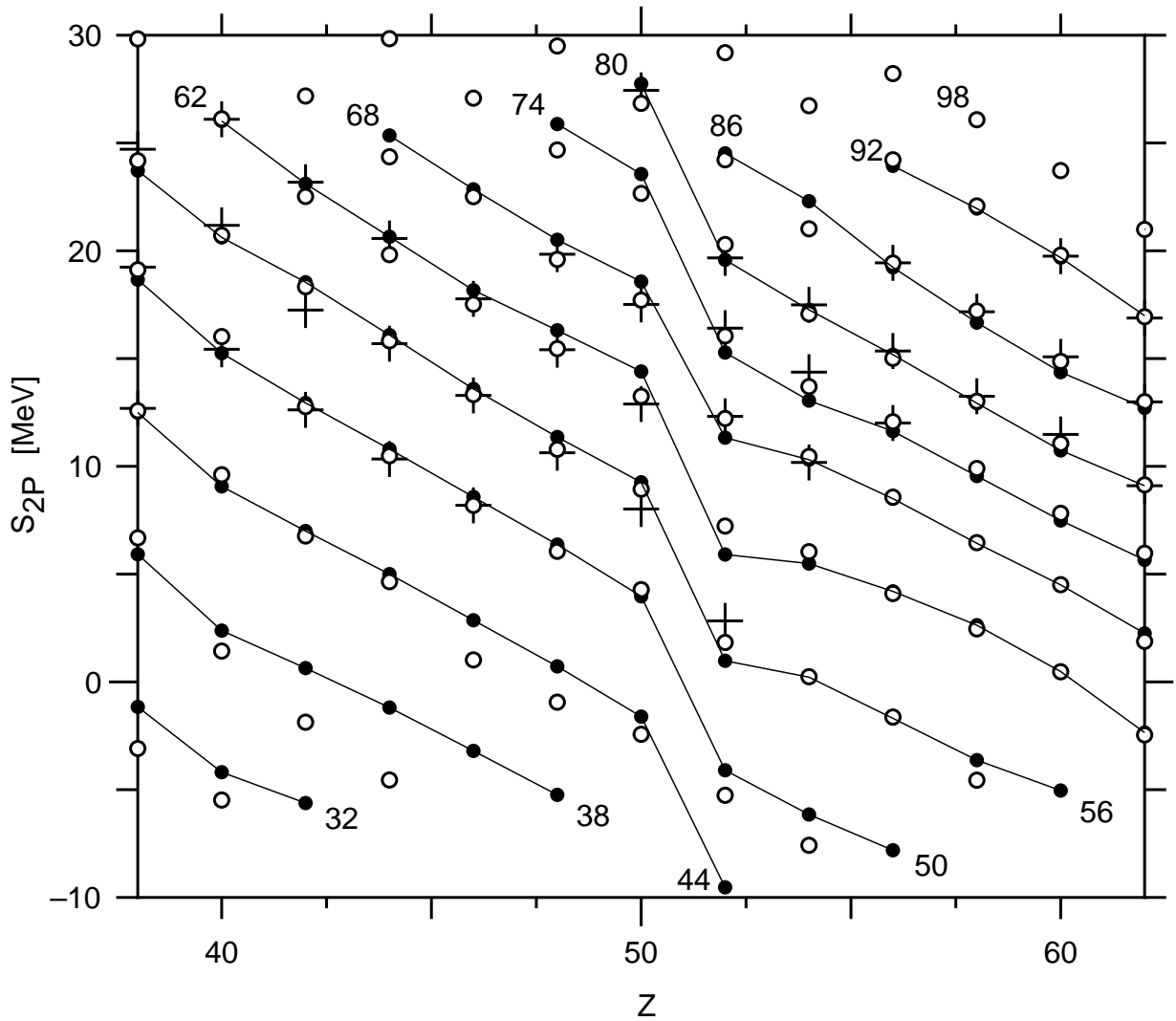
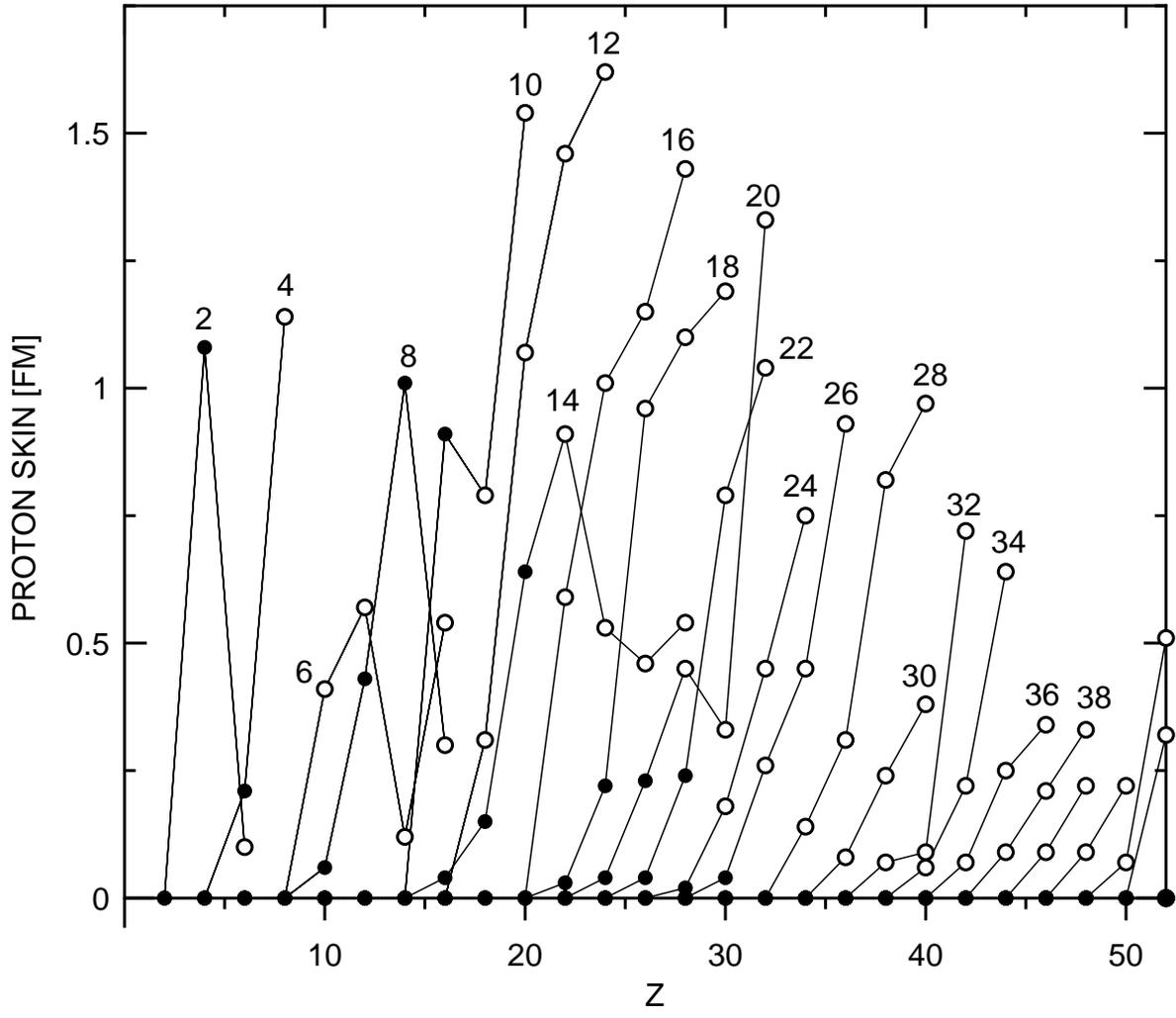
\end{document}